\documentclass[useAMS,usenatbib,usegraphicx]{mn2e}

\newcommand{\snd}{secondary}

\newcommand{\pri}{primary}
\newcommand{\rv}{radial velocity}
\newcommand{\rvs}{radial velocities}

\newcommand{\zdg}{$^{\circ}$}

\newcommand{\ks}{km s$^{-1}$}
\newcommand{\mbs}{maximum blueshift}
\newcommand{\mrs}{maximum redshift}
\newcommand{\wtd}{white dwarf}
\newcommand{\ha}{H$\alpha$}
\newcommand{\hb}{H$\beta$}
\newcommand{\hg}{H$\gamma$}
\newcommand{\heo}{HeI $\lambda$4471}

\newcommand{\el}{emission line}
\newcommand{\els}{emission lines}
\newcommand{\swc}{s-wave component}
\newcommand{\nsc}{narrow s-wave component}
\newcommand{\bbc}{broad-base component}

\usepackage{times}
\usepackage{mathptm}

\begin{document}


\title[Spectroscopic Study of the IP EX Hydrae in Quiescence]{Spectroscopic Observations of the Intermediate Polar EX Hydrae in Quiescence}
\author[N. Mhlahlo, D.A.H. Buckley, V.S. Dhillon, S.B. Potter, B. Warner and P.A. Woudt]{N. Mhlahlo$^{1,2}$\thanks{E-mail:
nceba@circinus.ast.uct.ac.za}, D.A.H. Buckley$^{2}$, V.S. Dhillon$^{3}$, S.B. Potter$^{2}$, B. Warner$^{1}$ and \newauthor P.A. Woudt$^{1}$  
 \\
$^{1}$Astronomy Department, University of Cape Town, Rondebosch 7700, Cape Town, South Africa \\
$^{2}$South African Astronomical Observatory, Observatory 7935, Cape Town, South Africa \\
$^{3}$Physics and Astronomy Department, University of Sheffield, Sheffield, S3 7RH, UK}


\maketitle

\label{firstpage}

\begin{abstract}
Results from spectroscopic observations of the Intermediate Polar (IP) EX Hya in quiescence during 1991 and 2001 are presented.  Spin-modulated radial velocities consistent with an outer disc origin were detected for the first time in an IP.  The spin pulsation was modulated with velocities near $\sim500-600$ \ks.  These velocities are consistent with those of material circulating at the outer edge of the accretion disc, suggesting corotation of the accretion curtain with material near the Roche lobe radius.
Furthermore, spin Doppler tomograms have revealed evidence of the accretion curtain emission extending from velocities of $\sim500$ \ks\ to $\sim1000$ \ks.  These findings have confirmed the theoretical model predictions of King \& Wynn (1999), Belle et al. (2002) and Norton et al. (2004) for EX Hya, which predict large accretion curtains that extend to a distance close to the Roche lobe radius in this system.  

Evidence for overflow stream of material falling onto the magnetosphere was observed, confirming the result of Belle et al. (2005) that disc overflow in EX Hya is present during quiescence as well as outburst. 

It appears that the \hb\ and \hg\ spin \rvs\ originated from the rotation of the funnel at the outer disc edge, while those of \ha\ were produced due to the flow of material along the field lines far from the \wtd\ (narrow component) and close to the \wtd\ (broad-base component), in agreement with the accretion curtain model. 

\end{abstract}


\section{Introduction}

EX Hya is an Intermediate Polar (IP), a sub-class of magnetic Cataclysmic Variable Stars (mCVs) where a late-type main sequence star transfers material to the magnetic white dwarf star as the two stars orbit each other under the influence of their mutual gravitation.  Unlike in Polars, another subclass of mCVs, where the white dwarf is in synchronous rotation with the binary rotation ($P_{spin}=P_{orb}$), the \wtd\ in an IP is in asynchronous rotation with the orbital motion of the system.  EX Hya, however, is nearer synchronism than the majority of IPs as it has a spin period ($\sim$67.03 min) which is about $2/3$ its orbital period (98.26 min) \citep{mum67,hel87}, and is one of only six out of thirty nine confirmed IPs with its orbital period below the 2-3 h CV period gap \citep{nor04}.  It has an inclination $i=78^{\circ} \pm 1^{\circ}$. 

Recent studies have shown that EX Hya does not conform to the traditional IP model \citep{kin99,wyn00,bel02,nor04,bel05}.  This system has a large $P_{spin}/P_{orb}$ ratio ($\sim0.68$) implying that it cannot be in the usual spin equilibrium rotation since most IPs have been shown to attain spin equilibrium near $P_{spin}/P_{orb} \sim0.1$ \citep{kin99,wyn00}.
This further implies that the corotation radius is far greater than the circularisation radius ($R_{co} \gg R_{cir}$) and that EX Hya cannot possess a Keplerian disc.  Systems with Keplerian discs are expected to have $R_{co}<R_{cir}$ and thus a smaller $P_{spin}/P_{orb}$. 
These factors have prompted theorists to suggest that the spin equilibrium state in EX Hya is determined by $R_{co} \sim b$, where $b$ is the distance to the inner Lagrangian point, $L_{1}$ \citep{kin99,wyn00,nor04}.  In this model the accretion curtains extend to near the $L_{1}$ point, and EX Hya resembles an asynchronous Polar where most of the material accretes via the stream \citep{kin99,wyn00} and via both the ring of material near the Roche lobe radius of the \pri\ and the stream \citep{nor04}, depending on the orbital and spin periods of the system, and the magnetic field strength.  In the later publication it was shown that material in EX Hya is fed from a ring of material at the outer edge of the Roche lobe, and that for the $P_{spin}/P_{orb}$ of EX Hya, this mode of accretion is preferred over stream-fed accretion.

\label{sec:obsred} 
In this work we present spectroscopic data of EX Hya in quiescence obtained from the SAAO in 1991 (just before EX Hya went into outburst, and a day or two after outburst) and in 2001.  Outburst data of 1991 will be discussed in a later publication.

\section{\bf Observations And Data Reduction}

\subsection{1991 Observations}
\begin{table}
\begin{center}
\small
\begin{tabular}{cccc}  \hline
\multicolumn{1}{c}{\textbf{Date}} &
\multicolumn{1}{c}{\textbf{HJD (start)}} &
\multicolumn{1}{c}{\textbf{Time}} &
\multicolumn{1}{c}{\textbf{Spectra}}    \\ \hline     
     
24-04-91 & 2448371.3884097 &  $3.28$  &  $90$  \\
25-04-91 & 2448372.3603850 &  $2.93$  &  $72$  \\
29-04-91 & 2448376.2402973 &  $7.00$  &  $100$   \\

 24-03-01 & 2451993.5427099  & $2.79$  &  $42$   \\  
 25-03-01 & 2451994.3581177 & $3.77$  &  $56$    \\
 25-03-01 & 2451994.5147196 & $3.96$  &  $66$   \\
 26-03-01 & 2451995.4360089 & $1.94$  &  $48$  \\
 26-03-01 & 2451995.5412087 &  $2.92$  & $50$   \\ \hline
\end{tabular}   
\end{center}
\caption{\small Table of spectroscopic observations during quiescence in 1991 and 2001.  The column \textbf{Date} denotes the date at the beginning of the observing night (before midnight), the column \textbf{Time} denotes the number of observing hours and \textbf{Spectra} the number of spectra obtained.} \label{tabo:observ}
\end{table} 

EX Hya was observed in April of 1991 by Buckley et al. (1991) using the SAAO 1.9-m telescope with the Reticon photon counting system (RPCS) detector on the Cassegrain spectrograph.  A grating with a resolution of 1200 mm$^{-1}$ was used and a wavelength range of 4000 - 5080 \AA~was covered at a spectral resolution of $\Delta \lambda \sim$1.2 \AA~and at a time resolution of 100 - 120 s.  The spectrograph slit width was 250 $\mu$m ($\sim$1.5 arcsecs).  Wavelength calibration exposures were taken using a CuAr arc lamp.  
Three nights of observations (24, 25 and 29 April 1991) were covered in quiescence and, in total, 262 spectra were obtained.  The observing log is given in Table~\ref{tabo:observ} together with the starting times of the observations.  

Following wavelength-calibration and sky-subtraction, the data were flux calibrated using the spectra of the standard star LTT3864.
\begin{figure}
\begin{center}
\includegraphics[width=80mm]{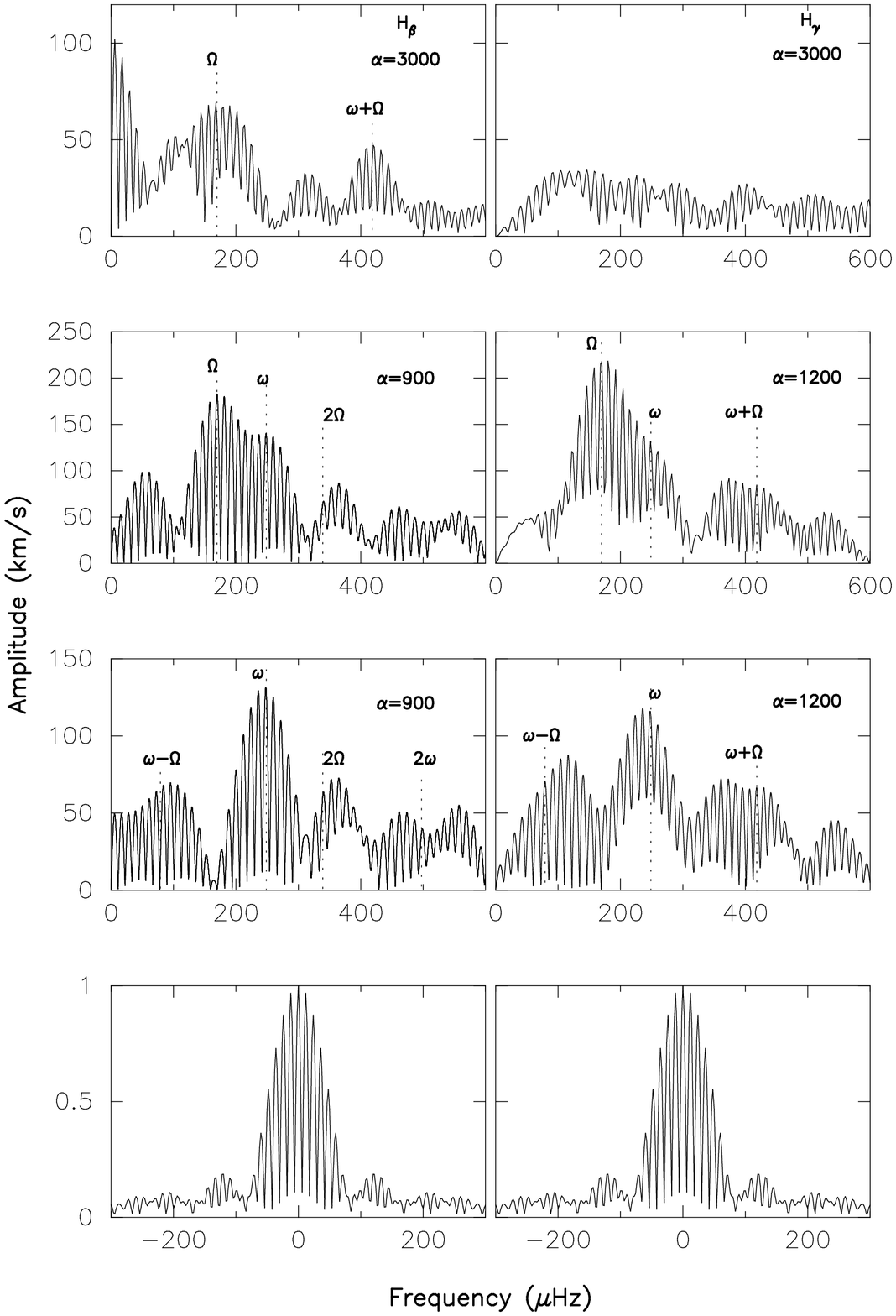}
\caption{\small Radial-velocity Fourier amplitude spectra from the 1991 combined data are shown for $\alpha=3000$ km s$^{-1}$ and 900 km s$^{-1}$ for the \hb\ line (top left panel and second left panel from the top) and for 3000 km s$^{-1}$ and 1200 km s$^{-1}$ for the \hg\ line (top right panel and second right panel from the top).  $\Omega$ denotes the orbital frequency of the system, 2$\Omega$ its first harmonic and $\omega+\Omega$ the upper orbital side band where $\omega$ is the spin frequency.  The data were prewhitened by the orbital frequency and are displayed in the third panels from the top.  Window spectra are plotted below the amplitude spectra (bottom panels).  }
\label{q:amplspec1}
\end{center}
\end{figure}

\subsection{2001 Observations}

The 2001 observations were obtained using the SITe CCD detector ($266\times1798$ pixels) on the Cassegrain spectrograph of the SAAO 1.9-m telescope.  A grating with a resolution of 1200 mm$^{-1}$  was used over the wavelength range 4200 - 5100 \AA~on the nights of the 25th and 26th April.  Another grating with a resolution of 1200 mm$^{-1}$ was used on the 24th, 25th and 26th April over the range 6300 - 7050 \AA.  The spectral resolution was $\sim$1.0 \AA~and a 1$\times2$ binning scheme was employed (i.e. binning by $2\times$ in the spatial direction).
The exposure times during observations were 60 s.
The observations covered the period 24 April - 26 April 2001 and, in total, 262 spectra were obtained (Table~\ref{tabo:observ}).  The extraction and reduction of the data were performed the standard way using the Image Reduction and Analysis Facility (IRAF)\footnote{IRAF is a software package for the reduction and analysis of astronomical data distributed by National Optical Astronomy Observatory (NOAO) which is operated by the Association of Universities for Research in Astronomy (AURA)} package and the spectra were flux calibrated using observations of the standard star LTT3218.

\section{The Radial Velocities}
\label{sec:radvels}

It is widely accepted that in a canonical CV, the high velocity emission line wings are formed in the inner parts of the accretion disc orbiting close to the \wtd\ and thus should reflect its orbital motion \citep{shaf83,shaf84,shaf85}.  In IPs, however, high velocity emission line wings are formed in the gas streaming towards the \wtd\ at high velocities \citep{hel87,fer93}.  
\begin{figure}
\begin{center}
\includegraphics[width=60mm]{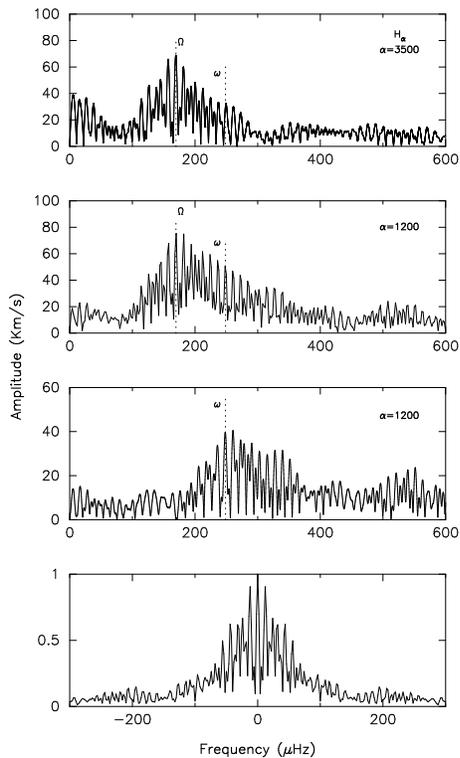}
\caption{\small Radial velocity amplitude spectra shown for the \ha\ line from 2001, for $\alpha=3500$ km s$^{-1}$ and 1200 km s$^{-1}$.  The vertical dashed line shows the expected position of the orbital and spin period peaks.  The data were prewhitened by $\Omega$ and are shown in the third panel from the top.  A window spectrum is shown at the bottom.}
\label{q:amplspec3}
\end{center}
\end{figure}
The radial velocities were determined by measuring the wings of the \hb\ and \hg\ emission lines from the 1991 data (24 and 25/04/91 - the data obtained on the 29th was not added since EX Hya had not fully recovered from outburst); and the wings of \ha, \hb\ and \hg\ emission lines from the 2001 data using the Gaussian Convolution Scheme (GCS, Schneider \& Young 1980; Shafter \& Szkody 1984; Shafter 1985).  The GCS method convolves each spectrum with two identical Gaussian, one in the red wing and one in the blue wing.  The separation between the two Gaussians is 2$\alpha$. 
Care was taken not to include regions far out in the wings where the continuum begins to dominate by choosing reasonable values of the width ($\sigma$) of the Gaussians and $\alpha$. 
Twelve standard Gaussian band-passes were used with $\alpha$ values ranging from 3500 to 100 km s$^{-1}$ and corresponding width values from 1200 to 100 km s$^{-1}$.
\subsection{Period Searches}
\label{sec:qampspec}
The \rvs\ were Fourier transformed using the Discrete Fourier Transform (DFT) algorithm to search for any periods in the data \citep{dee75,kur85}. The \hb\ and \hg\ amplitude spectra from 1991 (24 and 25 April) are shown in Figure~\ref{q:amplspec1} and the \ha\ amplitude spectra from 2001 are shown in Figure~\ref{q:amplspec3}.   
\begin{figure*}
\begin{center}
\includegraphics[width=175mm]{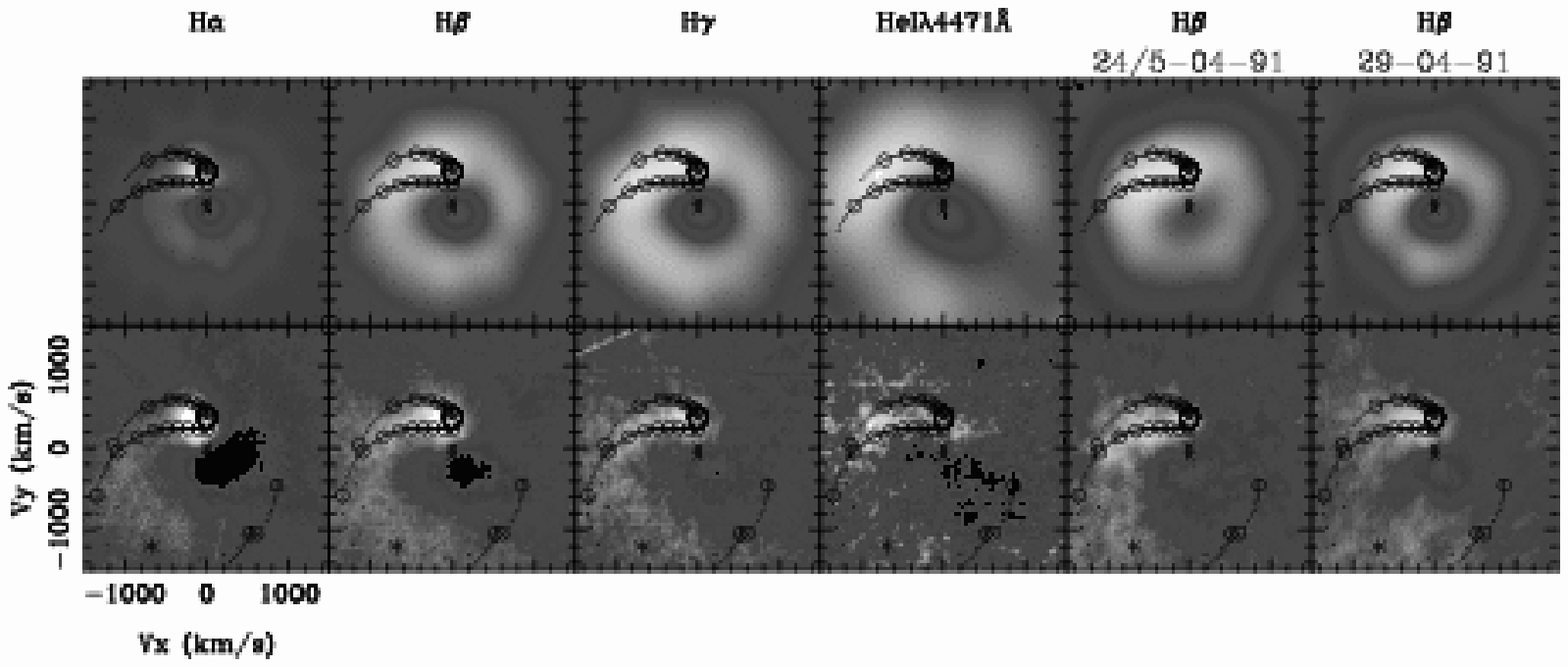}
\caption{\small The panels show the \ha, \hb, \hg\ and \heo\ orbital Doppler maps from 2001, and \hb\ tomograms of 1991, constructed using the Back-Projection Method with the application of a filter (top panels) and after subtracting the average of the line profile (bottom panels).  The positions of the Roche lobe and stream trajectories are shown (velocity amplitudes of $K_{1}=74$ km s$^{-1}$ and $K_{2}=360$ km s$^{-1}$ for the primary and secondary stars, respectively, were used).  The two curves with marked intervals represent the gas stream velocity (upper curve) and the Keplerian velocity along the stream (lower curve).  The circles on all tomograms represent 0.1 of the distance from the L$_{1}$ point to the \pri.  The three crosses are centres of mass of the \snd, system and \pri, from top-to-bottom.  The asterisk represents the velocity of closest approach.  All the maps are plotted on the same velocity scale.  The lookup table of this figure is such that the brightest emission features appear with decreasing intensity from yellow/green to light blue in the online edition, or white to grey in the printed edition.}
\label{q:habegatrl}
\end{center}
\end{figure*}

A prominent peak at a frequency corresponding to the 98-minute orbital frequency, $\Omega$, is observed in all the emission lines. 
Second in strength to the orbital frequency is the spin frequency, $\omega$, of the narrow s-wave component (NSC) ($\alpha=900, 1200$ \ks) (Figures~\ref{q:amplspec1} and~\ref{q:amplspec3}).  
The third panels from the top in Figures~\ref{q:amplspec1} and~\ref{q:amplspec3} show the data after prewhitening by $\Omega$.  Power at $\omega$ is clearly present.  

The spin frequency was not detected at high values of $\alpha$ ($\alpha=3000,3500$ km s$^{-1}$) where it is most expected (since at these velocities the material is quite close to the \wtd\ and its emission is expected to be modulated at the \wtd\ spin period).  It was also not present in \hb\ and \hg\ \rvs\ of 2001.
The amplitude of $\omega$ relative to $\Omega$ was found to be $\sim$81\% for \hb, $\sim$64\% for \hg\ and $\sim$18\% for \ha.  
\section{Orbital Variations of the Emission Lines}
\label{sec:qew}

The data were phase-binned on the orbital ephemeris of Hellier \& Sprouts (1992), 
\begin{equation}
T_{eclipse}=2437699.94179 + 0.068233846(4)E \label{eq:ephem},
\end{equation}
where $E$ is the number of orbital cycles and T is the time of mid-eclipse.  This ephemeris is defined by the zero point of mid-eclipse, where minimum intensity is at phase 0.0.  This means that for the \rvs, \mbs\ is perpendicular to the line of centres, at phase 0.75, meaning that spectroscopic phase zero occurs at the blue-to-red crossing of the \el\ \rv\ curve.  40 phase bins were used to produce the \rvs.  

\subsection{Orbital Tomograms and Trailed Spectra}
\label{q:odopmap}
\begin{figure}
\begin{center}
\includegraphics[width=85mm]{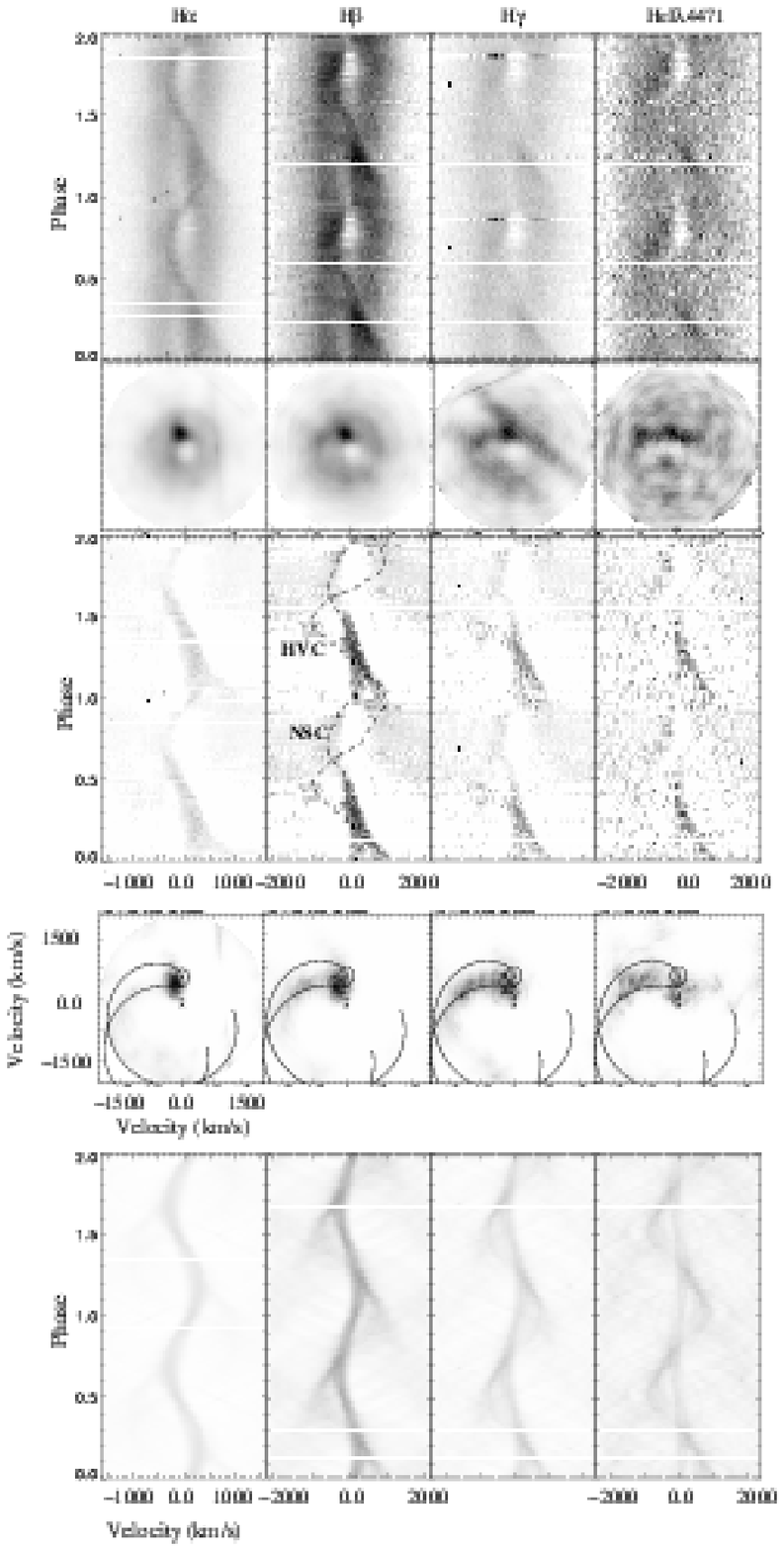}
\caption{\small 2001 \ha, \hb, \hg\ and \heo\ trailed spectra (top row of panels) and MEM orbital Doppler maps (second row of panels from the top) as well as the average-subtracted trailed spectra (third row of panels) are shown plotted on the same scale except for \ha\ panels.  The HVC and the NSC are indicated.  The fourth row shows the average-subtracted Doppler maps and the models plotted for $q=0.21$, $i=78^\circ$ and $M_{1}=0.50 M_{\odot}$.  The bottom panels are the reconstruction of the average-subtracted data.  The fourth and bottom panels are also plotted on the same scale except for \ha\ panels.  The lookup table of this figure is such that the brightest emission features appear with decreasing intensity from black to light grey.}
\label{q:habegaidltrl}
\end{center}
\end{figure}
The \ha, \hb\ and \hg\ Doppler tomograms were computed using the Back Projection Method (BPM) with the application of a filter, and the Maximum Entropy Method (MEM) \citep{mah88,mar88,hor91,spr98}.  The BPM Doppler maps are shown in Figure~\ref{q:habegatrl} and those constructed using MEM are shown in Figure~\ref{q:habegaidltrl}.  
Error in ephemerides (both orbital and spin) is sufficiently small to phase all of our data accurately on the orbital and spin cycles.

A velocity amplitude of the primary, $K_{1}=74\pm2$ km s$^{-1}$ from our \rv\ measurements and that of the \snd, $K_{2}=360\pm35$ km s$^{-1}$, taken from Vande Putte et al. (2003) and Beuermann et al. (2003), were used to fix the positions of the Roche lobe and the stream trajectories on the tomograms.  A \snd\ mass, $M_{2}= 0.10 \pm0.01$ M$_\odot$ for EX Hya was derived from the up-to-date secondary mass-period relation of Smith \& Dhillon (1998).  The mass of the primary, $M_{1}$, was then determined from the above values using $\frac{K_{1}}{K_{2}}=\frac{M_{2}}{M_{1}}$, and was found to be 0.50$\pm$0.05 M$_\odot$.   

The \hb, \hg\ and \heo\ Doppler tomograms (Figure~\ref{q:habegatrl} and~\ref{q:habegaidltrl}) show strong emission at the bright spot, some at the Roche lobe and the stream, and some from the disc.  Those of \ha\ also show strong bright spot emission but less or no emission from the stream.  Disc emission is diminished in \ha\ when compared to other \els, especially at higher velocities.  This is more obvious in the BPM tomogram.  The bright spot emission falls near the region (-100, 350) km s$^{-1}$.  Average-subtracted trailed spectra (Figure~\ref{q:habegaidltrl}) show the corresponding NSC.  
$\sim1\times10^{-11} {\rm ergs\hspace{0.1cm} cm^{-2} \hspace{0.1cm} s^{-1}}$ ($\sim$60-70\%) of the original line fluxes is contained in the average-subtracted profiles of \hb\ and \hg, and $\sim1\times10^{-12} {\rm ergs\hspace{0.1cm} cm^{-2}\hspace{0.1cm} s^{-1}}$ ($\sim$80-90\%) is contained in \ha\ and \heo.  This flux is mainly due to the bright spot and the stream.  The trailed spectra have been repeated over 2 cycles for clarity.  

The trailed spectra of 2001 have revealed two interesting features.  The first one is the asymmetry in the intensity of the s-wave (Figure~\ref{q:habegaidltrl}).  In \ha, the red wing of the NSC is brighter at $\phi_{98} \sim 0.1-0.3$ and seems to reach maximum brightness near $\phi_{98} \sim 0.25$, whereas the blue wing is dimmer in the range $\phi_{98} \sim 0.7-0.9$ and seems to reach minimum brightness near $\phi_{98} \sim 0.75$.
A similar effect is seen in the \hb\ and the \hg\ lines, and to a lesser extent in \heo.  
The second feature is redshifted emission extending from the NSC to high velocities ($\sim1000$ \ks) at early binary phases ($\phi_{98} \sim 0.0-0.2$).  

The reconstructed trailed spectra suggest that this latter feature is another s-wave, which we shall refer to as the high velocity component (HVC), crossing the NSC near $\phi_{98} \sim0.2-0.3$.  The Doppler tomograms show emission extending from the bright spot position, passing along the stream path, to the bottom-left quadrant at high velocities near 1000 \ks\ which is responsible for the HVC; most of this emission does not fall within the disc and gas stream velocities on the map, suggesting that there was small or no overlap of the stream component with the disc.  

The 1991 tomograms showed similar results.  It is worth noting that most of the disc emission in 1991 came from the outer disc than in 2001.

Even though the MEM and BPM pick out the same features, in the BP tomograms some features are more prominent than in the MEM tomograms while the reconstruction obtained using the MEM reproduces the observed data well.  

The advantage of BPM over MEM is that it is faster and it is easier to get a consistent set of maps of different emission lines (in terms of the apparent noise in the images).  For this reason both methods have been used.   

\section{\bf Spin Variations of the Emission lines}
\label{sec:svemq}

\subsection{The Spin Radial Velocity Curve}
 The \rvs\ were phase-folded using 30 bins on the quadratic spin ephemeris of Hellier \& Sproats (1992), where spin maximum was defined as $\phi_{67}=0$.
\begin{figure}
\begin{center}
\includegraphics[width=65mm]{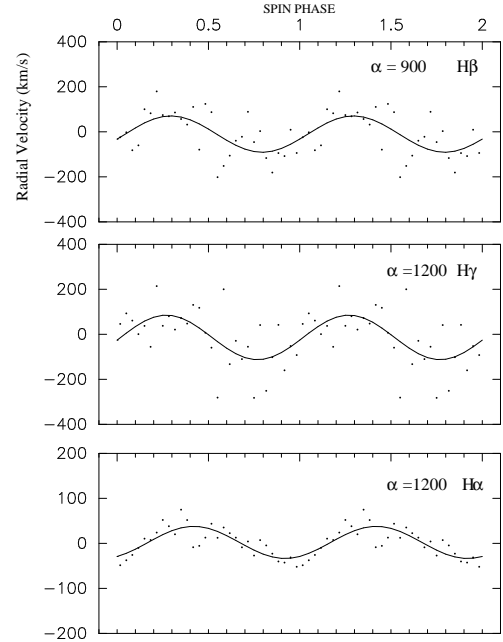}
\caption{\small The \hb\ (top panel), \hg\ (middle panel) and \ha\ (bottom panel) spin radial velocities of the narrow component from the 1991 combined data (\hb\ and \hg) and 2001 data (\ha).  The \rvs\ were prewhitened by the orbital frequency and phase-folded on the spin frequency using 30 bins and are shown plotted as a function of the spin phase.} 
\label{rv:hbg_245sppr}
\end{center}
\end{figure}
Figure~\ref{rv:hbg_245sppr} shows the variation of the \hb, \hg\ and \ha\ narrow components with $\omega$.  
Maximum blueshift is seen at $\phi_{67}=0.79$ for \hb\ and at $\phi_{67}=0.77$ for \hg.  Whereas for \ha, maximum blueshift is seen at $\phi_{67}=0.90$.  It should be noted that \ha\ and \hb\ / \hg\ have not been observed simultaneously and so both data sets probably sample the spin phases at different orbital phases.

\begin{figure}
\begin{center}
\includegraphics[width=70mm]{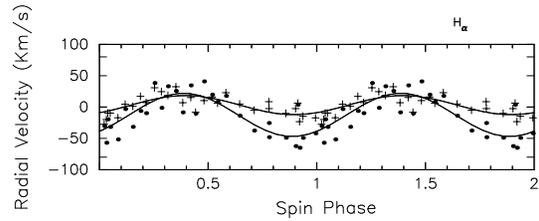}
\caption{\small The spin radial velocity curves of the \ha\ narrow (crosses) and broad (dots) components from 2001 (30 bins) plotted as a function of the spin phase.  The solid line represents a fit to the data.}
\label{rv:2457ha-s}
\end{center}
\end{figure}

Figure~\ref{rv:2457ha-s} shows the \ha\ narrow component ($\alpha=1200$ \ks) and the \bbc\ ($\alpha=3500$ \ks) overplotted.  The two components are in phase.  The radial velocity variation with the spin period of the \hb\ and \hg\ \bbc\ could not be detected, possibly due to velocity cancellation   We discuss this in Section~\ref{sec:oqdic}.  
\subsection{Spin Tomograms and Trailed Spectra}
\label{sec:spintomo}

Spin tomograms of EX Hya were constructed by \cite{hel99} but revealed little information.  Also, \cite{bel05} observed no coherent emission site/s on their tomograms folded on the spin phase.  

\begin{figure}
\begin{center}
\includegraphics[width=80mm]{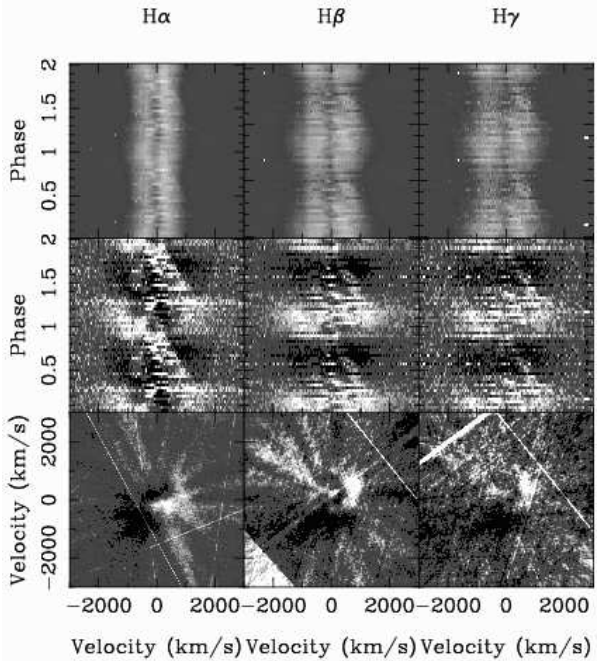}
\caption{\small The \ha, \hb, \hg\ and HeI $\lambda$4471 trailed spectra from 2001 folded on the spin period are shown at the top panels and the average-subtracted spectra are shown at the second panels.  Doppler maps constructed from the phase-invariant subtracted spectra are shown in the bottom panels.  The Doppler maps were constructed using the BPM and are shown on the same velocity scale with the trailed spectra.  The lookup table is as in Figure~\ref{q:habegatrl}.}
\label{q:spintrl}
\end{center}
\end{figure}

\begin{figure}
\begin{center}
\includegraphics[width=80mm]{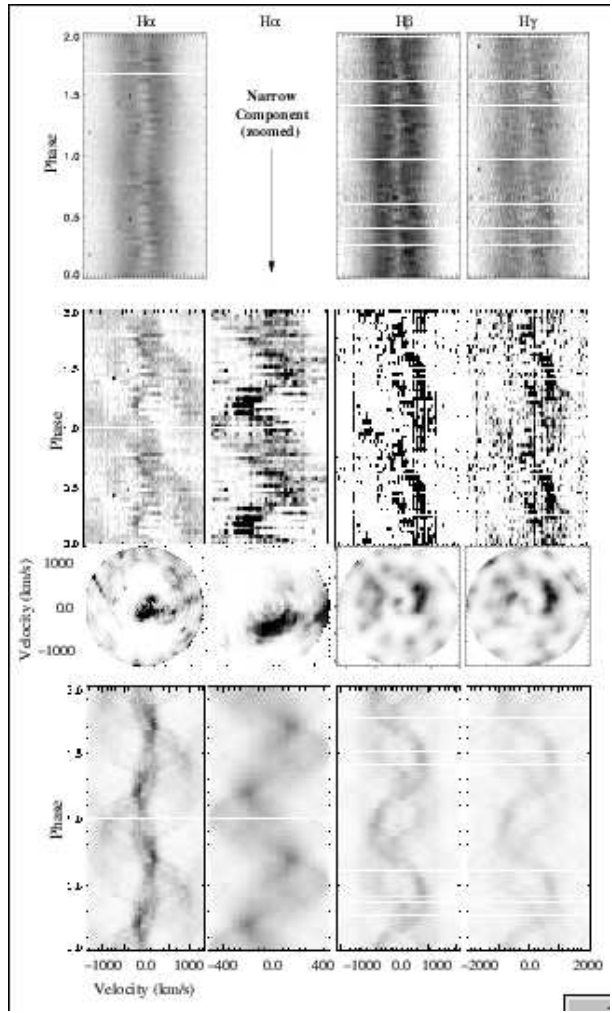}
\caption{\small \ha, \hb\ and \hg\ trailed spectra of 2001 folded on the spin period are shown in the top panels and the phase-invariant subtracted trailed spectra are shown in the second panels from the top.  MEM spin Doppler tomograms constructed from the phase-invariant subtracted spectra are shown in the third panels with the reconstructed spectra in the bottom panels.  The spin wave observed in the \ha\ phase-invariant subtracted trailed spectra, which was caused by the \ha\ narrow component, is shown expanded on a smaller velocity scale.  The first column of panels are plotted between -1500 \ks\ and 1500 \ks\ and the last two column are plotted between -2000 \ks\ and +2000 \ks.  The lookup table is as in Figure~\ref{q:habegaidltrl}.}
\label{q:habgspintrlmem}
\end{center}
\end{figure}

The \hb\ and the \hg\ BPM and MEM spin tomograms from 2001, however, have revealed a coherent emission site between V$_{x}\sim$ 500 km s$^{-1}$ and $\sim$ 1000 km s$^{-1}$ which is evidence of emission from the accretion curtains (Figures~\ref{q:spintrl} and \ref{q:habgspintrlmem}).  But it is a well known fact that since the spin period is $\sim \frac{2}{3}$ of the orbital period in EX Hya, orbital cycle variations do not smear out when folded on the spin phase but repeat every 3 spin cycles \citep{hel87}.  This is thought to be the origin of most of the structure in the \els\ at velocities $<1000$ \ks\ \citep{hel99}.  
To address this problem, phase-invariant subtraction is performed where emission that does not vary with the spin cycle is subtracted from the data.  This is achieved by measuring minimum flux at each wavelength and subtracting this value.  The results are shown in the second panels from the top of Figure~\ref{q:habgspintrlmem}.  It should be mentioned though, that even subtracting the invariant part of the line profiles does not guarantee that the influence of the orbital period variations has been completely removed.

A spin-wave (to differentiate it from the s-wave which is normally caused by the bright spot) in the \ha\ trailed spectra (Figure~\ref{q:habgspintrlmem}) was detected from the data after the phase-invariant subtraction was performed.  This is the first detection of modulation over the spin cycle in the optical emission line data of EX Hya.  This spin-wave can be seen in the trailed spectra before (but hard to see) and after subtraction of the phase-invariant line profile. 
The narrow peak component is responsible for this spin-wave which is shown expanded in the second column of panels in Figure~\ref{q:habgspintrlmem} (the narrow peak component was selected by hand over a velocity range of $\pm500$ \ks).  
The spin-wave shows maximum blueshift near phase 1.0 and maximum redshift near phase 0.5, and has an amplitude of $\sim$ 500 km s$^{-1}$.  The \ha\ tomogram shows corresponding emission near the ``3 o'clock'' position (blob of emission right at the edge of the map), around $\sim$ 500 km s$^{-1}$.  
DFTs show lower amplitude ($\sim40$ \ks\ for \ha\ and $\sim130-140$ \ks\ for \hb\ and \hg) probably due to dilution by stationary material.
Also, the \ha\ MEM tomogram shows stronger emission that peaks in a broad structure at lower velocities (V$_{x} \sim$ -200 km s$^{-1}$ -- $\sim$ +200 km s$^{-1}$ -- around the ``5-6 o'clock'' position).  Circular motion gives rise to low or zero radial velocities when the motion is perpendicular to the line of sight, and the emission seen around the ``5-6 o'clock'' position could not be from such velocities since it shows \mbs\ at $\phi_{67}\sim0.2-0.25$.  Similar emission was observed in a Polar and was thought to be due to material that has just been decelerated after having attached to the magnetic field lines \citep{sch05} (we discuss an alternative explanation in Section~\ref{sec:oqdic}).  The emission near the edge of the tomogram (at $\sim$ 500 km s$^{-1}$) shows \mbs\ at $\phi_{67}\sim1.0$ and therefore cannot be due to motion perpendicular to the line of sight either.

The \ha\ trailed spectra also show emission coupled to the spin-wave near $\phi_{67} \sim 0.1-0.6$ that extends to high velocities in the red, a similar situation to that seen the in orbital tomograms due to the HVC (Section~\ref{q:odopmap}).  The corresponding emission in the \ha\ tomograms extending to higher velocities in the red spectral region is not clear.  

Both the \hb\ and \hg\ phase-invariant subtracted trailed spectra show three weak-intensity spin-waves.  The most clearly visible of the three is phased with maximum redshift near $\phi_{67} \sim 0.3-0.4$, with an estimated velocity amplitude of $\sim 900$ km s$^{-1}$ and corresponds to the emission near the ``3 o'clock'' position in the tomogram (Figure~\ref{q:habgspintrlmem}).  The reconstructed trailed spectra reproduce the observed data.

The \hb\ and \hg\ trailed spectra in Figure~\ref{q:spintrl} seem to support these results.  The emission observed near the ``3 o'clock position'' in the tomograms has also been seen in other IPs such as AO Psc and FO Aqr \citep{hel99} and was interpreted as emanating from the upper accretion curtain.

The spin-waves are weak in intensity though, and more data are needed to support these results.

\section{Discussion of the Orbital and Spin Data}
\label{sec:oqdic}
The generally accepted model of EX Hya has the material leaving the \snd\ star through the L$_{1}$ point, passing via a stream of material which orbits about the \wtd, to form an accretion disc.  The magnetic field lines of the \wtd\ which form accretion curtains above and below the orbital plane channel the material from the disc, starting from the co-rotation radius ($R_{co}$) where the disc is truncated by the field lines, to the surface of the \wtd\ \citep{hel87,ros91}.

King \& Wynn (1999) challenged this model by arguing that systems with $P_{spin}/P_{orb} > 0.1$ cannot possess Keplerian discs since this implies $R_{co} \gg R_{cir}$.  They showed that the spin equilibrium state in EX Hya is determined by $R_{co} \sim b$, where $b$ is the distance to the $L_{1}$ point.  In this model the accretion curtains extend to near the $L_{1}$ point, and EX Hya resembles an asynchronous polar where most of the material accretes via the stream \citep{kin99,wyn00}.  

Belle et al. (2002) revised the model of EX Hya after they showed that their EUV data support the model of King \& Wynn (1999).  Their revised model suggested that the magnetic field in EX Hya forms a large accretion curtain extending to the outer edge of the Roche lobe causing: 
\begin{itemize}
\item{part or all of the non-Keplerian disc (hereafter the ring of material or the ring) to rotate with the \wtd},
\item{an extended bulge (later, Belle et al. (2005) showed that there was Vertically Extended Material (VEM) obscuring the s-wave emission during $\phi_{98}=0.57-0.87$, and evidence for overflowing stream accretion in EX Hya)}, and
\item{the ring of material to feel magnetic force at the regions of the ring close to the poles, causing the ring material at these locations to be controlled by the magnetic field, forming two chunks along the accretion ring that rotate with the \wtd.} 
\end{itemize}

Recently, Norton et al. (2004, 2004a) have shown that for systems with $P_{spin}/P_{orb} \sim0.72$, when the mass ratio is smaller at $q=0.2$, the material forms a ring near the edge of the \pri\ Roche lobe, from where accretion curtains funnel down to the \wtd\ surface, in agreement with King \& Wynn (1999) and Belle et al. (2002).  The material is fed from the ring (ring-fed accretion) and channeled along the magnetic field lines (when the angle between the \wtd\ spin axis and magnetic dipole axis is small i.e $< 30$\zdg, which is true for EX Hya).

The discussion by \cite{eis02} on the IR-UV flux distribution in EX Hya implies
a disc (isobaric and isothermal) with an outer radius of 1.6$\times10^{10}$ cm and a thickness of 2$\times10^{8}$ cm, and an assumed central hole of 6$\times10^{9}$ cm, but \cite{eis02} suggested that the structure could also be a ring with a larger inner radius, in line with the suggestion of \cite{kin99,bel02} and \cite{nor04}.  They found that the disc component contains about 1/6 of the total flux which is a bit more than expected from gravitational energy release at the inner radius, R$_{in} > 6\times10^{9}$ cm.  

Our spectroscopic data support both the model of Belle et al. (2002) and Norton et al. (2004) in which material from a ring, circling the \wtd\ and co-rotating with the magnetic field lines at the outer edge of the Roche lobe, is accreted by the \wtd.

The presence of the bright spot revealed by the trailed spectra, the DFTs of the \rvs\ and the Doppler maps (Figures~\ref{q:habegatrl} and \ref{q:habegaidltrl}) suggest the presence of a disc or ring of material extending to near the Roche lobe radius, around the \wtd.  When comparing the 1991 and 2001 tomograms for the \hb\ they appear to be in the same state or similar, given the fact that they are ten years apart.  It is reassuring that the fact that the two groups of lines have not been measured simultaneously is not a significant problem in the analysis.

More importantly, a spin pulse modulated at velocities consistent with those of the material circulating at the outer edge of the disc ($\sim500-600$ \ks) (Figures~\ref{q:amplspec1} and~\ref{q:amplspec3}) was detected and provides evidence for co-rotation of the extended accretion curtains with the ring material.  As discussed in Section~\ref{sec:spintomo}, these low radial velocities mentioned above were not caused by motion perpendicular to the line of sight near the \wtd, neither were they caused by velocity cancellation as will be shown later.  

\begin{figure}
\begin{center}
\includegraphics[width=60mm]{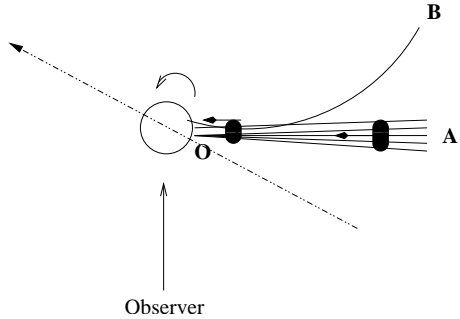}
\caption{\small A depiction of the regions where \ha\ was formed.  Both the narrow and broad base components fall along the same radial direction, OA, resulting in similar phase variation.}
\label{o:qexillus2e}
\end{center}
\end{figure}

A spin wave was detected in the spin-folded trailed spectra of \ha\ (Figure~\ref{q:habgspintrlmem}) with a velocity semi-amplitude of $\sim$ 500-600 km s$^{-1}$.  The spin wave shows maximum blue-shift near phase $\phi_{67}\sim1.0$ (when the upper magnetic pole is pointed away from the observer) and maximum redshift near phase $\phi_{67}\sim0.5$.  The \ha\ equivalent widths show maximum flux near $\phi_{67}\sim1.0$.  This picture is consistent with the accretion curtain model of IPs and is possible if accretion occurs via a disc/ring.  
The spin tomograms (Figures~\ref{q:spintrl} and~\ref{q:habgspintrlmem}) show evidence of the accretion curtain emission extending from $\sim 500$ \ks\ to high velocities ($\sim 1000$ \ks), suggesting that material is channeled along the field lines from the outer ring.  The \ha\ narrow and broad base components show similar phase variation, suggesting same position of maximum radial velocity as shown in Figures ~\ref{rv:2457ha-s} and \ref{o:qexillus2e} (line OA).  This indicates that material is channeled from the ring (at low velocities) to high velocities along the field lines.

A mass ratio of $q \sim$ 0.2 was measured from our data, and so the period ratio $P_{spin}/P_{orb}$ $\sim0.68$ is consistent with the ring accretion model of Norton et al. (2004).

Decreased prominence of the \nsc\ around $\phi_{98}=0.57-0.87$ (Figure~\ref{q:habegatrl} and~\ref{q:habegaidltrl}) was observed and suggests the presence of VEM at the outer edge of the ring of material obscuring the emission at these phases.  The presence of the overflow stream may be infered from this observation \citep{bel05}. 
But direct evidence comes from orbital Doppler tomograms which show an asymmetry in the emission, where more emission is observed from the \snd\ Roche lobe to the lower left quadrant than from the opposite side.  Average-subtracted orbital tomograms show this emission at higher velocities ($\sim$900-1000 \ks) (Figures~\ref{q:habegatrl} and~\ref{q:habegaidltrl}), and it corresponds to the HVC observed in the trailed spectra, which is modulated with a velocity semi-amplitude of $\sim 1000$ \ks.  
This HVC is reminiscent of that detected by Rosen et al. (1987) in the trailed spectra of the AM Her system V834 Cen.  Their HVC was blueshifted with a velocity of $\-900$ \ks and was said to be produced in the stream close to the \wtd.  The only difference is that there was no evidence of the HVC emission when it was expected to be seen redward of another component (medium-velocity component) in their data, whereas in EX Hya the evidence of the HVC emission is missing between $\phi_{98} \sim0.3-0.85$.  The HVC emission is maximally blueshifted at $\phi_{98} \sim0.3-0.4$.  This phasing is consistent with the expected phase of impact of a stream of material from the \snd\ with the disc or of the overflow stream material free-falling onto the magnetosphere of the \pri\ \citep{hel89}.  

Support for overflow stream is also provided by spin tomograms where emission is observed on the upper accretion curtain with velocities consistent with stream velocities.  This suggests that this emission site may also have resulted due to impact of overflow stream with the magnetosphere.   The resulting emission is receding from the observer at \mrs\ near $\phi_{67} \sim 0.4$ (Figure~\ref{q:habgspintrlmem}), in agreement with the accretion curtain model.  

The model of \cite{kin99} is not fully supported by our observations since it predicts direct accretion via a stream.  Our observations, however, fit the models of \cite{nor04} and \cite{bel02,bel05}.
There is evidence for strong \ha\ emission of the narrow s-wave component in the spin tomograms, centred around $\sim100$ \ks\ (Figure~\ref{q:habgspintrlmem}), that is not accounted for by these models.  This emission shows \mbs\ at phase $\phi_{67}\sim0.2$, suggesting that these are rotational velocities (or a combination of streaming and rotational velocities) of the antiphased motion of a source locked to the \wtd.  One possible explanation is that this emission comes from the opposite pole of the \wtd, at a radial distance of 6$\times10^{9}$ cm ($\sim$8R$_{wd}$).  \cite{sie89} found that the eclipsed optical source in EX Hya is centred at a radial distance of 1.5$\times10^{9}$ cm ($\sim$2R$_{wd}$), which is about four times closer to the \wtd\ compared to our result.  This could be the same emission region, but in our observations the emission is spread out, possibly due to the quality of the data, and this could account for the difference in the radial distance values quoted above.  But we cannot imagine a geometry where such low rotational velocities can dominate over streaming velocities along the field lines near the \wtd.  We therefore suggest that this is evidence for material that is diverted out of the orbital plane.  Since one of the assumptions of Doppler tomography is that everything lies on the plane, it is not possible to locate the exact position of this emission relative to the \wtd.  
\begin{figure*}
\begin{center}
\includegraphics[width=90mm]{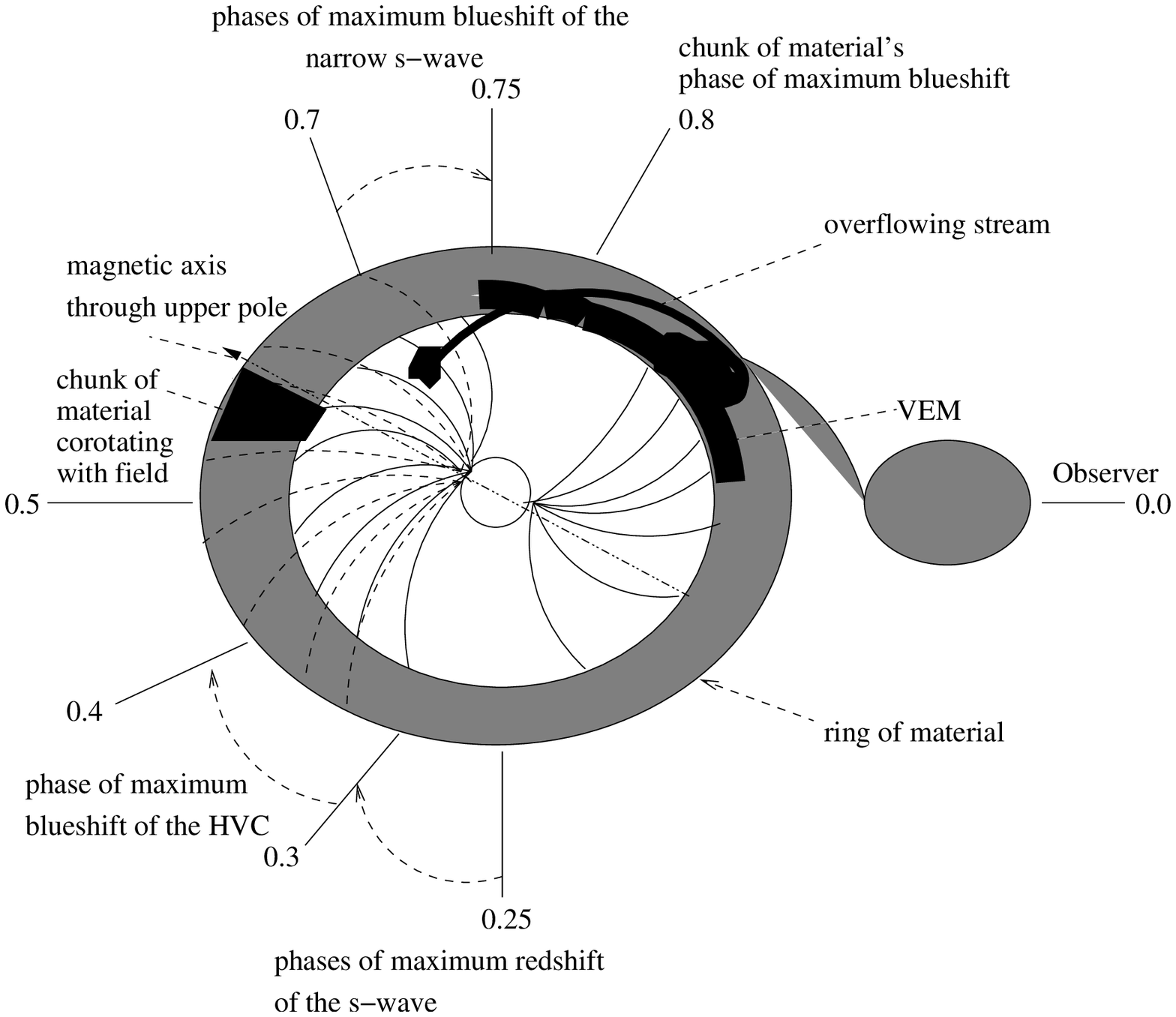}
\includegraphics[width=85mm]{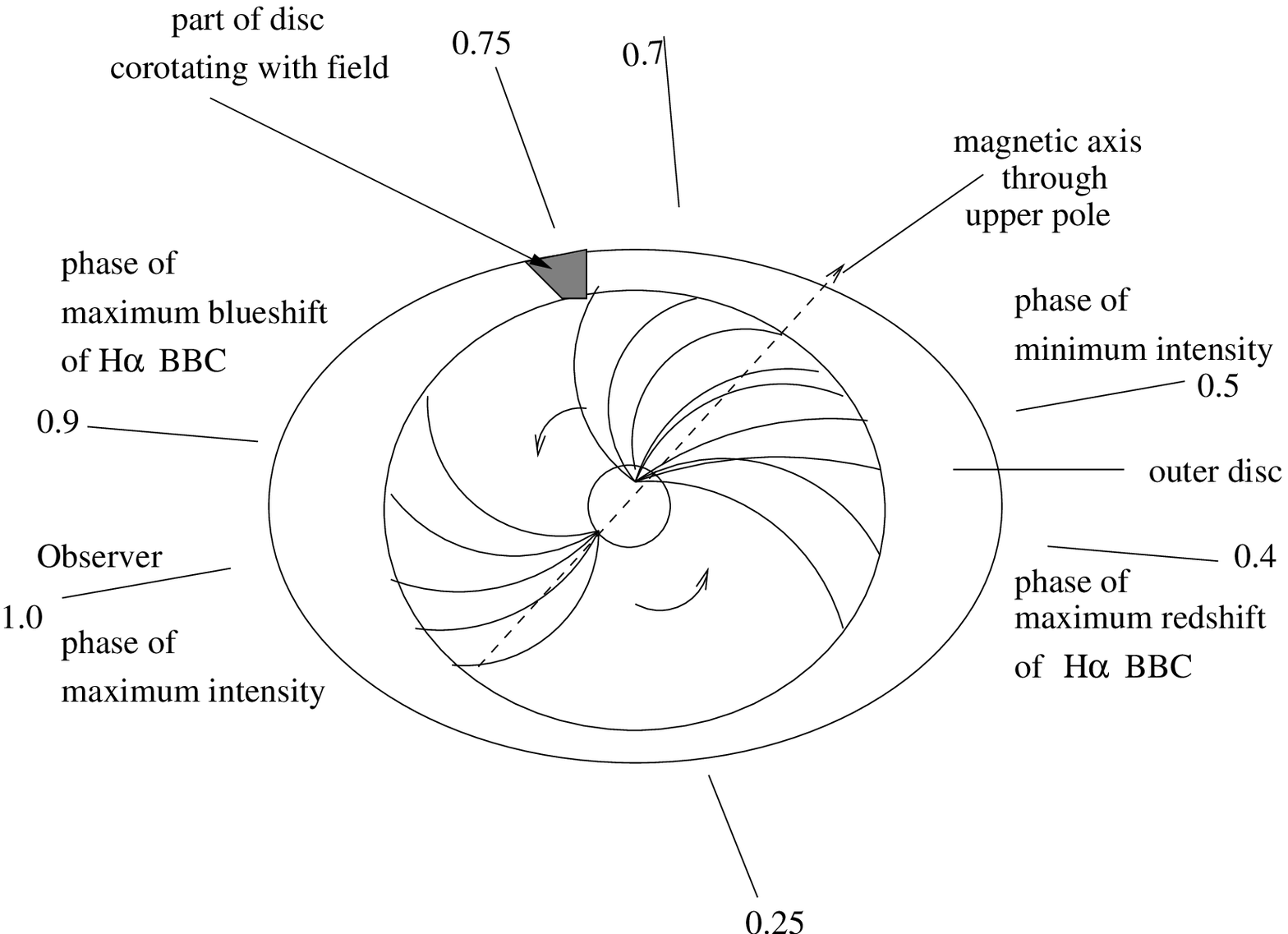}
\caption{\small A model of EX Hya in quiescence.  The figures are drawn over the orbital cycle (left) and spin cycle (right) and show the magnetosphere extending to the outer edge of the ring, and the chunk of material corotating with the field lines.  A verticaly extended material (VEM) is irradiated by the \wtd\ in its inner regions (left).}
\label{o:qexillus2b}
\end{center}
\end{figure*}
\subsection{White dwarf and \snd\ masses}

\cite{hel87} showed that maximum line widths of $\pm3500$ \ks\ constrain the mass of the \wtd, and a free-fall velocity of this magnitude could be achieved for \wtd\ masses greater than 0.48 M$_\odot$.  We found M$_{1}=0.50\pm0.05 M_\odot$, in good agreement with the results obtained from recent studies by \cite{hoo04,beu03} and \cite{put03}.

For the \snd, we derived M$_{2}=0.10 \pm0.01$ M$_\odot$ from the secondary mass-period relation of Smith \& Dhillon (1988), and this value agrees with that obtained by \cite{put03}.  \cite{beu03} and \cite{hoo04} find lower values for M$_{2}$ consistent with 0.09 M$_\odot$.  \cite{eis02} argues that for a \snd\ mass as low as 0.1 M$_\odot$ the secondary would have to be substantially expanded by $\sim$10\%.  

\subsection{The revised model of EX Hya}
We propose a model where one of the two chunks alluded to by Belle et al. (2002), which are formed by the magnetic pull along the accretion ring, co-rotates with the accretion curtains at the outer edge of the Roche lobe at $\sim$ 500-600 km s$^{-1}$, giving rise to the pulsation of emission at the spin period which we observe in our data, while the other is hidden by the accretion curtain below the ring of material. 
The resulting emission is maximally blueshifted near $\phi_{67} \sim0.8$ (Figure~\ref{rv:hbg_245sppr}).  In the accretion curtain model, at $\phi_{67} \sim0.5$ in the spin cycle, minimum flux (due to higher opacity) is observed when the upper accretion pole of the \wtd\ is pointed towards the observer \citep{hel87}, and so the phasing mentioned above is compatible with the motion of a rotating accretion funnel.  This is illustrated in Figure~\ref{o:qexillus2b}, where the position of the observer at pulse maximum is indicated, and the axis of the magnetic pole is shown.  The disruption of the disc by the magnetic field at the outer disc is illustrated and part of the disc co-rotating with the magnetosphere is shown.  
At a corotation radius, $R_{c} \sim$ \(b=a(0.500-0.227\log \frac{M_{2}}{M_{1}})\) ($\sim3\times10^{10}$ cm), the material is rotating at a velocity of $v^{2}=\frac{GM}{b}$ $\sim500$ \ks, in good agreement with the observations.
Also, a rotation velocity of $\sim600$ \ks\ was measured from the spectra and the radial distance from the star to the ring of material was found to be $\sim3\times10^{10}$ cm, which is similar to ${b}$, for a \wtd\ mass of 0.5 M$_{\odot}$ (Keplerian motion about the \wtd\ had to be assumed in these calculations).  At this radius, the accretion curtain is also rotating at a velocity of 2$\pi R_{co}$/P$_{spin}$ $\sim500$ \ks.  

$\sim6\times10^{-12}$ ergs cm$^{-2}$ s$^{-1}$ (64\% - integrated over one spin cycle) of the original line fluxes that is contained in the average-subtracted profile of \ha\ shows \rv\ variations with the spin period.  Assuming that \hb\ and \hg\ also show a similar flux variation (\hb\ and \hg\ spin tomograms also show a low-velocity s-wave but this result is not secured due to poor quality of data), the total line fluxes showing \rv\ variations with the spin period can be estimated to be $\sim2\times10^{-11}$ ergs cm$^{-2}$ s$^{-1}$ for the three \els.  This is $\sim$2/10 of the total disc flux \citep{eis02}, suggesting that only part of the ring corotates with the \wtd\ while the rest of the material may be involved in a near Keplerian motion (this is a rough comparison since the flux is integrated over one spin cycle for \ha, \hb\ and \hg\ in our data whereas \cite{eis02} derived their total flux values from one spectrum over the wavelength range $\lambda=912 - 24000$ \AA).

While some of the ring material co-rotates with the accretion curtains (i.e. remains in the disc rather than being immediately channeled along the field lines), some is channeled along the field lines at $\sim500$ \ks\ towards the \wtd.  There is also some material that overflows the ring and attaches onto the magnetic field lines.  
The overflow stream hits the magnetosphere, probably causing a second bright spot on the slowly rotating magnetosphere (Figure~\ref{o:qexillus2b}).  
The overflow stream is irradiated by the \wtd\ in its inner regions close to the \wtd\ (the regions facing the \wtd).  This results in the HVC emission being obscured at $\phi_{98} \sim0.4-0.9$, which are phases where the stream is viewed from behind-opposite the side facing the \wtd, hiding the irradiated inner regions.  HVC emission from the stream is blueshifted when that from the \nsc\ shows maximum redshift.  Near $\phi_{98}\sim0.25$ the two s-waves intersect, explaining the asymmetry in the brightness of the s-wave seen near $\phi_{98}\sim0.25$ (Figure~\ref{q:habegaidltrl}).  The overflow stream curls nearly behind the \wtd\ and it is truncated by the field when the upper magnetic pole is facing the stream.  

Ferrario \& Wickramasinghe (1993) and Ferrario, Wickramasinghe \& King (1993) showed that in IPs the accretion curtain below the orbital plane can contribute in the radial velocities of a system if it can be seen either through the central hole of the truncated disc, or from below the disc, or both.  This effect will result in velocity cancellation due to nearly equal quantities of material that are blueshifted and redshifted on the accretion curtains (Ferrario, Wickramasinghe \& King 1993).  

In EX Hya where the inclination is high (78\zdg) and the disruption radius is large ($\sim$ 40 R$_{WD}$, for a \wtd\ mass of 0.5 M$_{\odot}$) as proposed in Figure~\ref{o:qexillus2b}, it is clear that we see spin-varying emission from two opposite magnetic poles, producing a fairly symmetric structure in the spin-folded line profiles \citep{hel87,ros91}.  If emission from these opposite poles is cancelling out then the sum will have a much lower velocity.  This could explain the near zero and low amplitude of the \rv\ variation at the spin period of the \hb\ and \hg, and \ha\ ($\leq$ 40 \ks) \bbc, respectively (see also \cite{hel87} and Ferrario, Wickramasinghe \& King (1993)).

One could take this argument further by suggesting that the spin modulation we observe in our data at velocities near $\sim500$ \ks\ (Figures~\ref{q:amplspec1} and \ref{q:amplspec3}) is just the slight asymmetries between the two poles.  The resulting velocity could just be a measure of the degree to which the poles cancel their velocities near $\pm3500$ \ks (Coel Hellier; private communication).  This, however, cannot be the case for \hb\ and \hg\ since these two emission lines show motion that is consistent with that of a rotating object, suggesting that the line profiles are not dominated by the infall velocities at the two opposite accretion poles.  If they were produced close to the \wtd\ then maximum rotational velocity near $\pm3500$ \ks\ would be 2$\pi R/P_{spin} \sim$ 30 km s$^{-1}$, which is much smaller than $\sim500$ \ks.  However, rotional velocities close to the ring are $\sim500$ \ks.  
For \ha, however, we observe \mbs\ at $\phi_{67} \sim1.0$, and so velocity due to cancellation anywhere between 0 and $\pm3500$ \ks are expected, depending on how much the two poles cancel.  If both accretion curtains are still visible and symmetric at large radii (which is possible as suggested by Ferrario, Wickramasinghe \& King (1993) and our model), velocity cancellation will still result in smaller amplitudes than those of $\sim500$ \ks observed in our data.  This would then count against the argument above.  
Furthermore, \ha\ orbital Doppler tomograms show strong emission at the bright spot.  If our model is correct, the field lines should also attract this \ha\ dominated material, which is chanelled along the field lines, as already shown above.  The velocity of this material due to streaming motion near the outer ring is less than that of the \ha\ broad-base component close to the \wtd, as expected.
A strong constraint on our model is that the disruption radius of EX Hya has been shown to be at 5-9$\times10^{9}$ cm \citep{hel87,beu03} which implies a \wtd\ magnetic moment of $\mu \sim 7 \times10^{31}$ G cm$^{3}$.   For our model this would imply that the accretion curtains do not extend to near the Roche lobe radius.  The theoretical analysis of \cite{kin99} and \cite{wyn00}, however, has shown that equilibrium rotation is possible if the magnetic moment in EX Hya falls within the range of 10$^{33}\le\mu \le 10^{34}$ G cm$^{3}$.  These are comparable to weakest field AM Hers below the period gap, and that EX Hya could possess such magnetic moments is supported to a certain extent by the average-subtracted trailed spectra of EX Hya that are reminiscent of \els\ seen in some Polars, e.g. V834 Cen (as discussed above), EF Eri \citep{cra81, cow82}, QS Tel \citep{rom03} and VV Pup \citep{dia94}.
Furthermore, \cite{cum02} raised the possibility that the magnetic fields in IPs are buried by the material due to high accretion rates and so are not really as low as they appear.  The ring structure in EX Hya could imply higher accretion rates in EX Hya than previously thought since the capacity of the ring of material to store matter may be low when compared to that of a classical disc, resulting in the accretion of more material than in a classical disc case. 

\section{Summary}

Optical observations of EX Hya and the analysis have suggested that large accretion curtains extending to a distance close to the L$_{1}$ point exist in this system.  The DFTs and spin tomograms have for the first time provided evidence for corotation of the field lines with the ring material near the Roche lobe.  Also, tomography and the phasing of the spin waves have suggested that feeding by the accretion curtains of the material from the ring (ring-fed accretion) takes place.  
These findings support the models of Belle et al. (2002) and Norton et al. (2004) for EX Hya and the simulations done by Norton et al. (2004a) which have shown that for systems with the parameters of EX Hya, the accreting material forms a ring at the outer edge of the \pri\ Roche lobe, from where accretion curtains funnel down to the \wtd\ surface.  

Evidence for stream overflow accretion has been observed.  The HVC caused by the overflow stream disappeared at $\phi_{98}\sim0.4-0.9$ due to obscuration by the stream.  Obscuration of the NSC at $\phi_{98}\sim0.57-0.87$ suggested the presence of the VEM which was irradiated by the \wtd\ in its inner regions.

The \ha\ \bbc\ shows a radial velocity variation with the spin period whereas that of \hb\ and \hg\ could not be detected.  The low-amplitude velocity variations modulated at the spin period for \ha\ and for \hb\ and \hg\ is explained in terms of velocity cancellation effects.

We have provided an explanation for the asymmetry in the intensity of the narrow \swc\ seen in EX Hya trailed spectra in the optical.  The narrow \swc\ and the HVC cross at $\phi_{98}\sim0.25$, resulting in the asymmetry in brightness that we observe at these phases.  

The spin-folded trailed spectra are not of good quality and more data are needed to confirm these results.  

\section*{Acknowledgments}

{NM would like to acknowledge financial support from the Sainsbury/Linsbury Fellowship Trust and the University of Cape Town.  We would like to thank Kunegunda Belle, Coel Hellier and Andrew Norton for invaluable discussions and for their constructive comments.  We acknowledge use of D. O'Donoghue's and Tom Marsh's programs Eagle and Molly, respectively.}

\addcontentsline{toc}{chapter}{References}
\renewcommand{\bibname}{References}

\end{document}